\documentclass[10pt]{article}

\def\pmb#1{\setbox0=\hbox{#1}%
\kern-.025em\copy0\kern-\wd0
\kern.05em\copy0\kern-\wd0
\kern-.025em\raise.0433em\box0 }
\newcommand{\la}{\langle}
\newcommand{\ra}{\rangle}
\newcommand{\be}{\begin{equation}}
\newcommand{\ee}{\end{equation}}
\newcommand{\ba}{\begin{eqnarray}}
\newcommand{\ea}{\end{eqnarray}}
\newcommand{\om}{\omega}
\newcommand{\Om}{\Omega}

\newcommand{\pa}{\partial}
\newcommand{\br}{{\bf r}}
\newcommand{\bx}{\hat{\bf x}}

\newcommand{\bp}{\hat{\bf p}}

\newcommand{\bK}{\hat{\bf K}}
\newcommand{\hK}{{\hat{K}}}
\newcommand{\hC}{\hat{C}}
\newcommand{\hg}{\hat{g}}
\newcommand{\eps}{\epsilon}

\newcommand{\bE}{\hat{\bf E}}
\newcommand{\bF}{\hat{\bf F}}
\newcommand{\bH}{\hat{\bf H}}
\newcommand{\bA}{\hat{\bf A}}
\newcommand{\bD}{\hat{\bf D}}

\newcommand{\bP}{\hat{\bf P}}

\newcommand{\bk}{{\bf k}}

\newcommand{\bek}{{\bf e}_{{\bf k}\lambda}}
\newcommand{\ha}{\hat{a}}
\newcommand{\had}{\hat{a}^{\dag}}
\newcommand{\hb}{\hat{b}}
\newcommand{\hbd}{\hat{b}^{\dag}}
\newcommand{\hbb}{\hat{\bf b}}

\newcommand{\hB}{\hat{B}}
\newcommand{\hc}{\hat{c}}

\begin{document}
\title{Quantum Fields in a Dielectric: Langevin and Exact Diagonalization Approaches}
\author{F. S. S. Rosa, D. A. R. Dalvit, and P. W. Milonni\\
Theoretical Division, MS B213, Los Alamos National Laboratory, NM 87545, USA} 
\date{}
\maketitle
\vspace{5mm}

\section{Introduction}
Professor Erber has made important contributions to several areas of both pure and applied physics, making it easy to
identify topics about which one can write to celebrate his work; approaching such topics at his level of rigor and insight is far more difficult! His contributions to fundamental electromagnetic theory and quantum mechanics in particular include papers on electromagnetic energy density in dispersive media \cite{emenergy}, synchrotron-Cerenkov radiation \cite{sync}, radiation reaction \cite{rr}, and quantum jumps \cite{erberqm}, to cite only a few of those with which we are familiar. The first two papers cited, among others, deal with electromagnetic processes in dielectric media, and a small part of that subject will be addressed here. Specifically, this article is concerned with the quantized electromagnetic field in a dispersive and dissipative dielectric medium, and with the energy density in such a medium. Erber's work has also dealt with nonlinear processes in such media as well as in vacuum, but we will restrict ourselves here to linear and idealized, homogeneous media.

This is an important subject about which much has been written, although in most of the literature
it has been assumed that the medium can be assumed to be non-dissipative at field frequencies
of interest. The theory ignoring dissipation is not without value; it can be used to describe, for example,
the spontaneous emission by an atom in a host dielectric that does not absorb radiation at the atom's transition
frequency. But a dispersive medium cannot be non-absorbing at all frequencies. What happens, for instance,
if the medium in our example strongly absorbs radiation at the transition frequency? And what about
situations in which we cannot exclude any frequencies {\sl a priori} and therefore cannot ignore absorption? In the calculation of the van der Waals force between two neutral dielectric bodies, for example, all field frequencies can in principle contribute to the force. For two perfectly conducting parallel plates at zero temperature, similarly, 
Casimir \cite{casimir} discovered, as a consequence of the zero-point electromagnetic energy of every mode, that there is an attractive force per unit area between the plates. His original method involving changes in zero-point field energy was later extended to dielectrics by van Kampen {\sl et al.} \cite{vankampen} and others \cite{pwmbook}. As emphasized by Ginzburg \cite{ginzburg}, however, these theories invoking changes in zero-point energy ignore absorption:
 `` ... oddly enough there is no mention that they consider directly only transparent media" \cite{remark1}. An entirely different route, based on the calculation of the force via the stress tensor, was taken by Lifshitz \cite{lifshitz}; he accounts for absorption through the  fluctuation-dissipation relation between the quantum fluctuations of the polarization density and the imaginary (absorptive) part of the permittivity. 

The intent here is to derive, in probably the simplest way possible, expressions for the quantum electromagnetic field in a dispersive and dissipative dielectric medium, treating the medium as a continuum. The derivation might seem superfluous in the sense that correct expressions for the electric and magnetic fields in such a medium
are already available \cite{hutt}. However, the diagonalization procedure by which these expressions are obtained
is not easily applied to general, inhomogeneous media, whereas the ``Langevin approach" presented here can be
applied more or less straightforwardly when extended and formulated via Green functions \cite{buhm}. 

We begin in the following section with the simple model of an oscillator A coupled to a reservoir R of other oscillators, 
the R oscillators causing a damping of A described by a Langevin equation. 
 In Section \ref{sect3} we review the Fano procedure for the
diagonalization of the Hamiltonian of this system, and compare the diagonalization and Langevin-equation approaches
insofar as they describe the time evolution of A. In the limit of zero temperature, or at any finite 
temperature, the two descriptions are shown to be equivalent. In Section \ref{sect4} we generalize these
considerations, following the Langevin approach, to the model of a homogeneous dielectric medium in which each atom is treated as a harmonic oscillator coupled to its own reservoir. The Langevin forces acting on the atoms give rise to a
noise polarization determined by the reservoir operators, and the fluctuating electromagnetic field caused by this noise polarization can therefore be expressed in terms these operators. In similar fashion to the model of Sections
\ref{sect2} and \ref{sect3}, the quantized electric and magnetic fields obtained in this way have exactly
the same form as obtained by Fano diagonalization. It is shown explicitly in Section \ref{sect5} that the zero-point energy per mode of frequency $\om$ is $(1/2)\hbar\om$ regardless of whether there is absorption at that 
frequency \cite{rosa}.  


\section{An Oscillator and a Reservoir}\label{sect2}

Consider an oscillator A of frequency $\om_0$ coupled to a reservoir R of other oscillators, a well-studied model for 
dissipation in quantum theory. To make things as simple as possible we will assume that the A-R coupling
involves only energy-conserving processes, and choose this coupling such that the Hamiltonian is
\be
\hat{H}=\hbar\om_0\had\ha+\int_0^{\infty}d\om\hbar\om\hbd(\om)\hb(\om)+
\hbar\sqrt{\gamma/\pi}\int_0^{\infty}d\om[\had\hb(\om)+\hbd(\om)a],
\label{ham}
\ee
with $[\ha,\had]=1$, $[\hb(\om),\hb(\om')]=0$, $[\hb(\om),\hbd(\om')]=\delta(\om-\om')$. (We use the
circumflex to denote operators.) Because only energy-conserving processes are
included, our model is consistent with the so-called ``rotating-wave approximation" (RWA). 
The coupling we have chosen results in a frictional damping rate $\gamma$ in the time evolution of A, as follows. 

The Heisenberg equations of motion for $\ha$ and $\hb(\om)$ are
\be
\dot{\ha}=-i\om_0a-i\sqrt{\gamma/\pi}\int_0^{\infty}d\om\hb(\om),
\label{heis1}
\ee
\be
\dot{\hb}(\om)=-i\om\hb(\om)-i\sqrt{\gamma/\pi}\ha.
\label{form1}
\ee
Using the formal solution 
\be
\hb(\om,t)=\hb_0(\om)e^{-i\om t}-i\sqrt{\gamma/\pi}\int_0^{t}dt'\ha(t')e^{i\om(t'-t)}
\ee
of equation (\ref{form1}) in equation (\ref{heis1}), and defining $\hb_0(\om)\equiv\hb(\om,0)$, we obtain 
\ba
\dot{\ha}(t)+i\om_0\ha(t)&+&(\gamma/\pi)\int_0^{\infty}d\om\int_0^tdt'\ha(t')e^{i\om(t'-t)} 
=-i\sqrt{\gamma/\pi}\int_0^{\infty}d\om\hb_0(\om)e^{-i\om t},
\ea
the operator on the right-hand side being a quantum Langevin force.
We solve this equation for ``steady state" ($\gamma t\gg 1$) by first writing
\be
\ha(t)=\int_0^{\infty}d\Om A(\Om)\hb_0(\Om)e^{-i\Om t}.
\ee
Then, using the approximation
\be
\int_0^tdt'e^{i(\om-\Om)(t'-t)}\cong \pi\delta(\om-\Om)-iP{1\over\om-\Om}
\label{pp}
\ee
for times $t$ such that $\Om t\gg 1$ for frequencies $\Om$ that make a significant contribution to the
time evolution of $\ha(t)$, we obtain
\be
A(\Om)={\sqrt{\gamma/\pi}\over\Om-\om_0+\Delta(\Om)+i\gamma}.
\ee
The frequency shift
\be
\Delta(\Om)={\gamma\over\pi}P\int_0^{\infty}{d\om\over\om-\Om}
\label{deldef}
\ee
obviously diverges in our model, and so the upper limit of integration must be appropriately cut off; for
our purposes there is no need to explicitly indicate any cutoff. Then our solution for $\ha(t)$ is 
\be
\ha(t)=\sqrt{{\gamma\over\pi}}\int_0^{\infty}d\Om{\hb_0(\Om)e^{-i\Om t}\over\Om-\om_0+\Delta(\Om)+i\gamma}.
\label{soln1}
\ee
Note that, under the assumption that $\om_0\gg\gamma$, consistent with the RWA, we have
\be
[\ha(t),\had(t)]={\gamma\over\pi}\int_0^{\infty}{d\om\over(\om-\om_0)^2+\gamma^2}\cong 1,
\label{rwacond}
\ee
as required for the validity of the RWA.  


\section{Fano Diagonalization}\label{sect3}

A method of diagonalizing a Hamiltonian for coupled oscillators, used many years ago by Fano \cite{fano},
has been employed in seminal work by  Huttner and Barnett \cite{hutt} to obtain expressions for the quantized 
fields in a {\sl dissipative} dielectric medium. Here we briefly review the method in the case of the model 
Hamiltonian (\ref{ham}), and compare it to the approach of the preceding section.

We define an operator
\be
\hB(\Om)=\alpha(\Om)\ha+\int_0^{\infty}d\om\beta(\Om,\om)\hb(\om)
\label{Bdef}
\ee
that we require to satisfy 
\be
[\hB(\Om),\hB^{\dag}(\Om')]=\delta(\Om-\Om'), \ \ \ [\hB(\Om),\hB(\Om')]=0, 
\label{comm}
\ee
and also require that the Hamiltonian (\ref{ham}) takes the diagonal form
\be
\hat{H}=\int_0^{\infty}d\Om\hbar\Om\hB^{\dag}(\Om)\hB(\Om).
\ee
From $[\hB(\Om),\hat{H}]=\hbar\Om\hB(\Om)$ and the definition (\ref{Bdef}) we deduce equations relating
the coefficients $\alpha(\Om)$ and $\beta(\Om,\om)$:
\be
(\om_0-\Om)\alpha(\Om)=-\sqrt{\gamma\over\pi}\int_0^{\infty}d\om\beta(\Om,\om),
\label{alphaeq}
\ee
\be
\int_0^{\infty}d\om\left[(\om-\Om)\beta(\Om,\om)+\sqrt{\gamma\over\pi}\alpha(\Om)\right]\hb(\om)=0.
\ee
The last equation leads us to write 
\be
\beta(\Om,\om)=\alpha(\Om)f(\Om)\delta(\om-\Om)-\sqrt{\gamma\over\pi}{\alpha(\Om)\over\om-\Om},
\ee
and it follows from (\ref{alphaeq}) that
\be
f(\Om)=\sqrt{\pi\over\gamma}[\Om-\om_0+\Delta(\Om)],
\ee
i.e.,
\be
\beta(\Om,\om)=\sqrt{\pi\over\gamma}\alpha(\Om)[\Om-\om_0+\Delta(\Om)]\delta(\om-\Om)-
\sqrt{\gamma\over\pi}{\alpha(\Om)\over\om-\Om},
\label{beta}
\ee
where $\Delta(\Om)$ is defined by (\ref{deldef}). 

To determine $\alpha(\Om)$ we impose the requirement that the commutation relations (\ref{comm}) be satisfied. From the
commutation relations stated earlier for the $\ha$ and $\hb$ operators we obtain
\ba
[\hB(\Om),\hB^{\dag}(\Om')]&=&\alpha(\Om)\alpha^*(\Om')
+ \int_0^{\infty}d\om\int_0^{\infty}d\om'\beta(\Om,\om)
\beta^*(\Om',\om')\delta(\om-\om')
\ea
or, from (\ref{beta}) and some straightforward algebra,
\ba
[\hB(\Om),\hB^{\dag}(\Om')]&=&{\pi\over\gamma}\alpha(\Om)\alpha^*(\Om')
\times \bigg{\{}
[\Om-\om_0+\Delta(\Om)]^2\delta(\Om-\Om') \nonumber \\
&& \hspace{-30mm} + {\gamma\over\pi}{\Delta(\Om')-\Delta(\Om)\over\Om'-\Om} 
\mbox{}+{\gamma^2\over\pi^2}P\int_0^{\infty}{d\om\over\om-\Om}P\int_0^{\infty}{d\om'\over\om'-\Om'}\delta(\om-\om')
\bigg{\}}.
\ea
Using
\be
P{1\over\om-\Om}={1\over\om-\Om+i\eps}+i\pi\delta(\om-\Om) \ \ \ (\eps\rightarrow 0^+)
\ee
and partial fractions we obtain
\be
[\hB(\Om),\hB^{\dag}(\Om')]={\pi\over\gamma}\alpha(\Om)\alpha^*(\Om')\left(\gamma^2+[\Om-\om_0+\Delta(\Om)]^2\right)
\delta(\Om-\Om').
\ee
Therefore we can satisfy (\ref{comm}) by taking
\be
\alpha(\Om)={\sqrt{\gamma/\pi}\over\Om-\om_0+\Delta(\Om)-i\gamma}.
\label{alpha}
\ee
Then
\be
\ha(t)=\int_0^{\infty}d\Om\alpha^*(\Om)\hB(\Om,t)=
\sqrt{{\gamma\over\pi}}\int_0^{\infty}d\Om{\hB(\Om,0)e^{-i\Om t}\over\Om-\om_0+\Delta(\Om)+i\gamma},
\label{soln2}
\ee
since $\hB(\Om,t)=\hB(\Om,0)e^{-i\Om t}$.

The expressions (\ref{soln1}) and (\ref{soln2}) obtained respectively in the Langevin and Fano approaches look 
formally the same in the sense that $[\hB(\Om,0),\hB^{\dag}(\Om',0)]=
[\hb_0(\Om),\hbd_0(\Om')]=\delta(\Om-\Om')$ and $[\hB(\Om,0),\hB(\Om',0)]=[\hb_0(\Om),\hb_0(\Om')]=0$. They
differ in that $\hB(\Om,0)$ in (\ref{soln2}) is a linear combination of 
A and R operators, whereas only R operators determine $\ha(t)$ in (\ref{soln1}). Suppose, however, that
at $t=0$ the A and R oscillators are all in their ground states. This state $|\Psi\ra$ is 
the {\sl exact} ground state of the coupled A-R system in the RWA:
\be
\hB^{\dag}(\Om,t)\hB(\Om,t)|\Psi\ra=0.
\ee  
In this case the properties of A derived from (\ref{soln1}) are trivially equivalent to those 
obtained from (\ref{soln2}). If the system is not initially in an eigenstate of $\hB^{\dag}\hB$,
it will nevertheless approach after a time $\gg\gamma^{-1}$ an equilibrium state for which the long-term solution
(\ref{soln1}) for $\ha(t)$ is applicable, i.e., transient effects associated with $\ha(0)$ at some initial
time $t=0$ ultimately play no role in the evolution of A. Equilibrium values of correlation functions involving products of the $\hB$ operators are determined solely by the reservoir operators; in thermal equilibrium, for example,
$\la\hB^{\dag}(\Om,t)\hB(\Om',t)\ra=\la\hbd(\Om,t)\hb(\Om',t)\ra=
[\exp(\hbar\Om/k_BT)-1]^{-1}\delta(\Om-\Om')$ and it follows from either (\ref{soln1}) or (\ref{soln2})
that $\la\had(t)\ha(t)\ra=[\exp(\hbar\om_0/k_BT)-1]^{-1}$ when we invoke the condition $\om_0\gg\gamma$ as in
(\ref{rwacond}). In other words, in any state of equilibrium the solutions (\ref{soln1}) and (\ref{soln2})
provide equivalent descriptions of A. This equivalence holds more generally beyond the RWA (see below).


\section{Fields in a Dielectric Continuum}\label{sect4}

Aside from the need to introduce oscillator strengths in order to obtain correct numerical results, we can model
a dielectric medium in which atoms remain with high probability in their ground states as a collection of
harmonic oscillators. We take each oscillator ``atom" to have a mass $m$ and a natural frequency $\om_0$. We
assume each of these material oscillators is coupled to a 
reservoir of other harmonic oscillators responsible for the damping of its oscillations and line broadening. For the Hamiltonian, including the electromagnetic field and its (electric-dipole) coupling to the material oscillators, we write
\ba
\hat{H}&=&{1\over 8\pi}\int d{\bf r}(\bE^2+\bH^2)+\sum_j \left( {1\over 2m}[\bp_j-{e\over c}\bA(\br_j)]^2 +{1\over 2}m\om_0^2\bx_j^2 \right) \nonumber \\
&& +\int_0^{\infty}d\om\hbar\om\sum_j\left[\hbb^{\dag}_j(\om)\cdot\hbb_j(\om)+{1\over 2}\right]  
- i\int_0^{\infty}d\om \Lambda(\om)\sum_j\bx_j\cdot[\hbb_j(\om)-\hbb_j^{\dag}(\om)] .
\label{eq1} 
\ea
The first two terms are the Hamiltonian operators for the electromagnetic field, the material oscillators (atoms),
and their coupling via the (Coulomb-gauge) vector potential $\bA(\br_j)$, $\br_j$ being the position of the $j$th
atom. The third and fourth terms are respectively the Hamiltonian operators for the reservoir oscillators and their 
interaction with the atoms. The reservoir oscillators satisfy the commutation relations
\be
[\hat{b}_{i\mu}(\om),\hat{b}_{j\nu}^{\dag}(\om')]=\delta_{ij}\delta_{\mu\nu}\delta(\om-\om') , \ \ \ 
[\hat{b}_{i\mu}(\om),\hat{b}_{j\nu}(\om')]=0  ,
\label{eq2}
\ee
where we use Greek letters to denote Cartesian components of vectors. We choose the atom-reservoir coupling constant to be
\be
\Lambda(\om)=\left({m\hbar\gamma\om\over\pi}\right)^{1/2}.
\label{eq3}
\ee
Then, as shown below, each atom's oscillations are damped at the rate $\gamma$. Note that no rotating-wave
 approximation is made in writing (\ref{eq1}). The operators $\bA$ and $\bE$ satisfy the usual canonical
 commutation relations for the electromagnetic field.

From (\ref{eq2}) and $[\hat{x}_{i\mu},\hat{p}_{j\nu}]=i\hbar\delta_{ij}\delta_{\mu\nu}$ we obtain the Heisenberg equations of motion
\be
{\ddot{\bx}}_j+\om_0^2\bx_j={e\over m}\bE(\br_j)+{i\over m}\int_0^{\infty}d\om \Lambda(\om)[\hbb_j(\om,t)-\hbb_j^{\dag}(\om,t)] ,
\label{eq4}
\ee
\be
{\dot{\hbb}}_j(\om,t)=-i\om \hbb_j(\om,t)+{1\over\hbar}\Lambda(\om)\bx_j \ .
\label{eq5}
\ee
Using the formal solution of (\ref{eq5}) in (\ref{eq4}), we write
\ba
{\ddot{\bx}}_j+\om_0^2\bx_j&=&{e\over m}\bE(\br_j)+{1\over m}\bF_{Lj}(t) 
+ {i\over m\hbar}\int_0^{\infty}d\om \Lambda^2(\om)
\int_0^tdt'\hat{\bf x}_j(t')[2i\sin\om(t'-t)],
\label{eq6}
\ea
where the Langevin force operator $\bF_{Lj}(t)$ acting on the $j$th atom is 
\be
\bF_{Lj}(t)=i\int_0^{\infty}d\om \Lambda(\om)[\hbb_j(\om,0)e^{-i\om t}-\hbb_j^{\dag}(\om,0)e^{i\om t}] .
\label{lang}
\ee
The third term on the right-hand side of (\ref{eq6}) is
\ba
-{2\over m\hbar}\int_0^{\infty}d\om \Lambda^2(\om)\int_0^tdt'\bx_j(t')\sin\om(t'-t) 
&=&-{2\gamma\over\pi}\int_0^tdt'\bx_j(t') \nonumber \\
\int_0^{\infty}d\om\om\sin\om(t'-t) 
={2\gamma}\int_0^tdt'\bx_j(t'){\partial\over\partial t'}\delta(t'-t)
&=&-\gamma{\dot{\bx}}_j(t) \ .
\label{damp}
\ea
We ignore a divergent frequency shift which, as in the model considered in the preceding sections, can
be made finite by introducing a form factor or a high-frequency cutoff; the (finite) shift can be assumed to be 
contained in the definition of $\om_0$. Equation (\ref{eq6}) then has the form of a quantum Langevin equation:
\be
{\ddot{\bx}}_j+\gamma{\dot{\bx}}_j+\om_0^2\bx_j={e\over m}\bE(\br_j)+{1\over m}\bF_{Lj}(t)  .
\label{eq7}
\ee

In the absence of coupling to the electromagnetic field we have, for times $t\gg\gamma^{-1}$,
\be
\bp_j(t)=m{\dot{\bx}}_j(t)=\int_0^{\infty}d\om\om \Lambda^2(\om)\left[{\hbb_j(\om)e^{-i\om t}\over\om_0^2-\om^2-i\gamma\om} \right.
\left.+{\hbb_j^{\dag}(\om)e^{i\om t}\over\om_0^2-\om^2+i\gamma\om}\right] .
\label{eq8}
\ee
(We now write $\hbb_j(\om)$ in place of $\hbb_j(\om,0)$.) Similarly, using (\ref{eq2}), we obtain 
\ba
[\hat{x}_{i\mu}(t),\hat{p}_{j\nu}(t')] &=&\delta_{ij}\delta_{\mu\nu}{2i\hbar\gamma\over\pi}\int_0^{\infty}
{d\om\om^2\cos\om(t'-t)\over(\om_0^2-\om^2)^2
+\gamma^2\om^2} \nonumber \\
&= & i\hbar\delta_{ij}\delta_{\mu\nu}\left[\cos\om_1(t'-t)-{\gamma\over 2\om_1}\sin\om_1|t'-t|\right]e^{-\gamma|t'-t|/2} ,
\label{eq9}
\ea
where $\om_1\equiv[\om_0^2-\gamma^2/4]^{1/2}$. The canonical commutation relation $[\hat{x}_{i\mu}(t),\hat{p}_{j\nu}(t)]=i\hbar\delta_{ij}\delta_{\mu\nu}$ is therefore preserved in the coupling of each atom to its reservoir. 

Since we are working in the Heisenberg picture, expectation values are over the initial state of the
coupled system of oscillators. If we assume that the reservoir is in an initial state of thermal equilibrium at temperature $T$, then
\ba
\la \hat{b}^{\dag}_{i\mu}(\om)\hat{b}_{j\nu}(\om')\ra = \la \hat{b}_{i\mu}(\om)\hat{b}^{\dag}_{j\nu}(\om')\ra-
\delta_{ij}\delta_{\mu\nu}\delta(\om-\om') 
={1\over e^{\hbar\om/k_BT}-1}\delta_{ij}\delta_{\mu\nu}\delta(\om-\om').
\label{btherm}
\ea

The Heisenberg equations of motion for the electric and magnetic fields that follow from the Hamiltonian (\ref{eq1}) 
and the canonical commutation relations for the field operators have exactly the same form as their classical (Maxwell) counterparts:
\begin{eqnarray}
\label{MaxEqs}
&&\nabla \times \hat{\bf E} = - \frac{1}{c}\frac{\partial\hat{\bf B}}{\partial t} , \nonumber \\
&&\nabla \times \hat{\bf H} = \frac{4 \pi}{c} \hat{{\bf J}} + \frac{1}{c}\frac{\partial\hat{\bf E}}{\partial t}.
\end{eqnarray}
For a charge-free medium, furthermore, $\nabla \cdot \hat{\bf B} = \nabla \cdot \hat{\bf D} =  0$, 
where
\ba
&&\hat{\bf D} =  \hat{\bf E} + 4\pi\hat{\bf P}, \nonumber \\
&&\hat{\bf J}({\bf r},t) = \frac{\partial\hat{\bf P}({\bf r},t)}{\partial t}, \nonumber \\
&&\hat{\bf P}({\bf r},t)= e\sum_j\bx_j(t)\delta^3(\br-\br_j),
\ea
with $\hat{\bf B} =  \hat{\bf H}$ in our model. 

It is convenient to work in the frequency domain and write
\ba
\bE(\br,t)&=&\int_0^{\infty}d\om[\bE(\br,\om)e^{-i\om t}+\bE^{\dag}(\br,\om)e^{i\om t}]  , \nonumber \\
\bH(\br,t)&=&\int_0^{\infty}d\om[\bH(\br,\om)e^{-i\om t}+\bH^{\dag}(\br,\om)e^{i\om t}]  , \nonumber \\
\bP(\br,t)&=&\int_0^{\infty}d\om[\bP(\br,\om)e^{-i\om t}+\bP^{\dag}(\br,\om)e^{i\om t}]  .
\label{eq11}
\ea
The Fourier transform of the polarization density may be written as
\ba
\bP(\br,\om)&=&e\sum_j\bx_j(\om)\delta^3(\br-\br_j), 
\label{eq12a}\\
\bx_j(t)&=&\int_0^{\infty}d\om[\bx_j(\om)e^{-i\om t}+\bx_j^{\dag}(\om)e^{i\om t}],
\label{eq13}
\ea
and it follows from (\ref{eq7}) that
\ba
\bP(\br,\om)&=&{e^2/m\over\om_0^2-\om^2-i\gamma\om}\sum_j\bE(\br_j,\om)\delta^3(\br-\br_j) 
\mbox{}+{ie/m\over \om_0^2-\om^2-i\gamma\om}\Lambda(\om)\sum_j\hbb_j(\om)\delta^3(\br-\br_j) \nonumber \\
&\rightarrow&{Ne^2/m\over\om_0^2-\om^2-i\gamma\om}\bE(\br,\om) 
+{iNe/m\over \om_0^2-\om^2-i\gamma\om}\Lambda_c(\om)\hbb(\br,\om) 
\label{eq15}
\ea
in the approximation in which we assume the atoms are continuously distributed with a density $N$ and we 
define $\Lambda_c(\omega) =  \sqrt{\rho_m \hbar\gamma\omega / \pi}$, with $\rho_m = m/N$.

From Maxwell's equations and (\ref{eq15}) we obtain
\be
\nabla^2\bE(\br,\om)+{\om^2\over c^2}\eps(\om)\bE(\br,\om)=-{\om^2\over c^2}\bK(\br,\om) ,
\label{eq17}
\ee
where the complex permittivity is defined as
\be
\eps(\om)=1-{4\pi Ne^2/m\over\om^2-\om_0^2 + i\gamma\om}\equiv 1- {\om_p^2\over\om^2-\om_0^2 + i\gamma\om} 
=\eps_R(\om)+i\eps_I(\om).
\label{eq18}
\ee
We have also defined the ``noise polarization" at frequency $\om$:
\be
\bK(\br,\om)={4\pi iNe/m\over\om_0^2-\om^2-i\gamma\om}\Lambda(\om)\hbb(\br,\om) .
\label{noisepol}
\ee
This noise polarization obviously stems from the Langevin force $\bF_{Lj}(t)$ in the quantum Langevin 
equation (\ref{eq7}). Its principal properties for our purposes are the thermal equilibrium expectation values
\ba 
\la \hK_{\mu}(\br,\om)\ra &=& \la \hK^{\dag}_{\mu}(\br,\om)\ra = 0 , \nonumber \\
\la \hK_{\mu}(\br,\om)\hK_{\nu}(\br',\om')\ra &=& \la \hK^{\dag}_{\mu}(\br,\om)\hK^{\dag}_{\nu}(\br',\om')\ra = 0 ,
\ea
and
\be
\la\hK^{\dag}_{\mu}(\br,\om)\hK_{\nu}(\br',\om')\ra=4\hbar\eps_I(\om)\delta_{\mu\nu}\delta(\om-\om')\delta^3(\br-\br')
{1\over e^{\hbar\om/k_BT}-1},
\label{prop1}
\ee
\be
\la\hK_{\mu}(\br,\om)\hK^{\dag}_{\nu}(\br',\om')\ra=4\hbar\eps_I(\om)\delta_{\mu\nu}\delta(\om-\om')\delta^3(\br-\br')
\left[ {1\over e^{\hbar\om/k_BT}-1}+1 \right] ,
\label{prop2}
\ee
all of which follow from (\ref{btherm}) and $\la \hat{b}_{i\mu}(\om)\hat{b}_{j\nu}(\om')\ra=0$. Equations (\ref{prop1})
and (\ref{prop2}) are the well-known fluctuation-dissipation relations between the
correlation functions of a noise polarization and the imaginary part of the dielectric function \cite{lifshitz,remark2}.

Next we define operators $\hg_{\lambda}(\bk,\om)$ by writing
\be
\hat{\bf K}(\br,\om)=\int d^3k\sum_{\lambda=1,2}\hg_{\lambda}(\bk,\om)\bek e^{i\bk\cdot\br}.
\label{eqk}
\ee
Since $\nabla\cdot\hat{\bf K}(\br,\om)=0$ we can choose the vectors $\bek$ such 
that $\bk\cdot\bek=0$, $\bek\cdot{\bf e}_{\bk\lambda'}=0$, $\lambda=1,2$; we also take the $\bek$ to be real. Then
\ba
\hg_{\lambda}(\bk,\om)&=&\left({1\over 2\pi}\right)^3\int d^3r\,\hat{\bf K}(\br,\om)
\cdot\bek e^{-i\bk\cdot\br} 
\equiv  \left({1\over 2\pi}\right)^3\int d^3r\hK_{\lambda}(\br,\om) e^{-i\bk\cdot\br} ,
\ea
and Eqs. (\ref{noisepol}) and (\ref{eq2}) imply the commutation relation 
\be
[\hg_{\lambda}(\bk,\om),\hg^{\dag}_{\lambda'}(\bk',\om') ] ={\hbar\over 2\pi^3}\eps_I(\om)\delta_{\lambda\lambda'}
\delta(\om-\om')\delta^3(\bk-\bk') .
\ee
We also define operators 
\be
\hc_{\lambda}(\bk,\om)\equiv [\hbar\eps_I(\om)/2\pi^3]^{-1/2}\hg_{\lambda}(\bk,\om) 
\label{eqkk}
\ee
satisfying 
\be
[\hc_{\lambda}(\bk,\om),\hc_{\lambda'}^{\dag}(\bk',\om')]=\delta_{\lambda\lambda'}\delta(\om-\om')\delta^3(\bk-\bk').
\label{eqC}
\ee

Finally an expression for the quantized electric field follows from (\ref{eq11}), (\ref{eq17}), (\ref{eqk}), and 
(\ref{eqkk}):
\ba
\bE(\br,t)= \int d^3k\sum_{\lambda}\int_0^{\infty}d\om\sqrt{\hbar\eps_I(\om)/2\pi^3}{\om^2/c^2\over k^2-
\eps(\om)\om^2/c^2}\hc_{\lambda}(\bk,\om)\bek 
e^{-i(\om t-\bk\cdot\br)}+{\rm h.c.} 
\label{eq33}
\ea
From $\nabla\times\bE=-(1/c)\pa{\hat{\bf B}}/\pa t$ we also obtain
\ba
\bH(\br,t)=i\int d^3k\sum_{\lambda}\int_0^{\infty}d\om\sqrt{\hbar\eps_I(\om)/2\pi^3}
\mbox{}{\om/c\over k^2-\eps(\om)\om^2/c^2}\hc_{\lambda}(\bk,\om)
\left( \bk \times \bek\right) e^{-i(\om t-\bk\cdot\br)} +{\rm h.c.} 
\label{eq34}
\ea

These expressions have the same form as the corresponding ones obtained by Huttner and Barnett \cite{hutt} by Fano diagonalization of the entire system of coupled harmonic oscillators (EM field, dielectric oscillators, and bath oscillators). Their equations for the quantized
electric and magnetic fields, however, involve annihilation and creation operators $\hC_{\lambda}(\bk,\om)$ and $\hC^{\dag}_{\lambda}(\bk,\om)$ for the exactly diagonalized Hamiltonian, instead of the reservoir annihilation 
and creation operators $\hc_{\lambda}(\bk,\om)$ and $\hc^{\dag}_{\lambda}(\bk,\om)$ appearing in our expressions
(\ref{eq33}) and (\ref{eq34}). Their diagonalized Hamiltonian, including the zero-point energy, is 
\be
H=\int d^3k\sum_{\lambda}\int_0^{\infty}d\om\hbar\om\left[\hC^{\dag}_{\lambda}(\bk,\om)\hC_{\lambda}(\bk,\om)+1/2\right].
\label{diagham}
\ee
The situation here parallels that for the simple model employed in Sections \ref{sect2} and \ref{sect3}, except that
no rotating-wave approximation has been made, and that one deals with three coupled subsystems instead of two coupled
subsystems: {\sl we again arrive at results by a straightforward ``Langevin" approach that
are equivalent to those obtained by diagonalizing the complete Hamiltonian}. Equations (\ref{eq33}) and (\ref{eq34})
are analogous to equation (\ref{soln1}) obtained in the Langevin approach to the single oscillator coupled to a reservoir,
whereas the Huttner-Barnett equations for the fields are analogous to equation (\ref{soln2}) obtained by
exact diagonalization. As in the model of Sections \ref{sect2} and \ref{sect3}, results such as (\ref{eq33}) and 
(\ref{eq34}) obtained by the Langevin approach will reproduce those obtained by exact diagonalization for
dielectric media in thermal equilibrium. To illustrate this we show in the next section that the {\sl total} zero-point
energy appearing in (\ref{diagham}) follows exactly from our approach; the calculation also sheds light on
some of the physics involved, and in particular on the role of the Langevin forces in maintaining equilibrium.


\section{Energy Density}\label{sect5}

We consider now the total energy density of the system of dielectric atoms, their reservoirs, and the electromagnetic
field, focusing for simplicity on the limit of zero temperature. We start from Poynting's theorem in the conventional notation, using the symmetrized Poynting operator
$\hat{\bf S}=(c/8\pi)[\bE\times\bH-\bH\times\bE]$ and taking expectation values
over the initial state of the system consisting of the field, the dielectric atoms, and the reservoir:
\be
\label{Poynting3}
\oint\la\hat{\bf S}\ra\cdot{\bf n}da=-{1\over 8\pi}\int\la\bE\cdot{\pa\bD\over\pa t}+{\pa\bD\over\pa t}
\cdot\bE\ra dV
-{1\over 8\pi}\int\la\bH\cdot{\pa\bH\over\pa t}+{\pa\bH\over\pa t}\cdot\bH\ra dV.
\ee
The left-hand side gives the energy flux through a closed surface $S$ and, given that we are assuming thermal 
equilibrium, must vanish. We identify the rate of change of the expectation value of the total energy density $W$ as
\be
{\pa W\over\pa t}={1\over 8\pi}\la\bE\cdot{\pa\bD\over\pa t}+{\pa\bD\over\pa t}\cdot\bE\ra 
+{1\over 8\pi}{\pa\over\pa t}\la\bH^2\ra ,
\ee
and the assumption of thermal equilibrium implies that this must also vanish. In the case of interest  $\bD=\bE+4\pi\bP_{\eps}+\bK$, where $\bP_{\eps}$ is the part of the
polarization giving rise to the dielectric permittivity $\eps(\om)$ and $\bK$ is the noise polarization defined by
(\ref{noisepol}). Thus $\bD=\bD_{\eps}+\bK$ and
\be
{\pa W\over\pa t}={\pa W_1\over\pa t}+{\pa W_2\over\pa t},
\ee
where 
\be
{\pa W_1\over\pa t}={1\over 8\pi}\la\bE\cdot{\pa\bD_{\eps}\over\pa t}+{\pa\bD_{\eps}\over\pa t}\cdot\bE\ra
+{1\over 8\pi}{\pa\over\pa t}\la\bH^2\ra
\ee
and
\be
\label{W2}
{\pa W_2\over\pa t}={1\over 8\pi}\la \bE\cdot{\pa\bK\over\pa t}+{\pa\bK\over\pa t}\cdot\bE\ra.
\ee

Before proceeding with the calculation of $W$ we note the following identity expressing conservation of energy:
\ba
{\pa W\over \pa t}&=&\la{\pa\over\pa t}\sum_j\left[{1\over 2}m{\dot{\bx}}_j^2+{1\over 2}m\om_0^2{\bx_j}^2\right]\delta^3(\br-\br_j)
+{1\over 4\pi}{\pa\over\pa t}\left[\bE^2+\bH^2\right]\nonumber \\
&&\mbox{}+\sum_j[2\gamma({1\over 2}m{\dot{\bx}}_j^2)-{\dot{\bx}}_j\cdot{\bf F}_{Lj}]\delta^3(\br-\br_j)\ra.
\label{phys}
\ea
The first term is the rate of change of the energy density (kinetic plus potential) of the oscillators 
of the dielectric, while the second term is the rate of change of the energy density of the
electromagnetic field. If there were no dissipation ($\gamma=0$ and therefore ${\bf F}_{Lj}=0$),
the third term on the right would vanish, and $W$ would be just the matter-plus-field energy density. 
The third term gives the rate of change of the energy density in the reservoirs; $2\gamma\sum_j({1\over 2}m{\dot{\bx}}_j^2)\delta^3(\br-\br_j)$ is the rate of increase of energy density in the reservoir, equal to
the rate at which the energy density of the dielectric oscillators decreases due to their coupling 
to their reservoirs, while $\sum_j{\dot{\bx}}_j\cdot{\bf F}_{Lj}\delta^3(\br-\br_j)$ is the rate of work per unit volume done by the Langevin forces of the reservoirs on the dielectric oscillators. 

Using (\ref{eq11}) and
\be
{\pa\bD_{\eps}\over\pa t}=-i\int_0^{\infty}d\om\om[\eps(\om)\bE(\br,\om)e^{-i\om t}
- \eps^*(\om)\bE^{\dag}(\br,\om)e^{+i\om t}],
\ee
and integrating over $t$, we obtain
\ba
W_1(\br,t)={1\over 8\pi}\sum_{\lambda} \int_0^{\infty}d\om'\int_0^{\infty}d\om
\frac{\om'\eps^*(\om')-\om\eps(\om)}{\om'-\om}
\la\bE_{\lambda}(\br,\om)\cdot\bE^{\dag}_{\lambda}(\br,\om')\ra 
e^{-i(\om-\om')t} \mbox{}+{1\over 8\pi}\la\bH^2(\br,t)\ra,
\label{w11}
\ea
since the zero-temperature expectation value 
$\la\bE^{\dag}_{\lambda}(\br,\om)\cdot\bE_{\lambda'}(\br,\om')\ra=0$ and
 $\la\bE_{\lambda}(\br,\om)\cdot\bE^{\dag}_{\lambda'}(\br,\om')\ra=0$ unless $\lambda=\lambda'$
and $\om=\om'$. It is convenient to rewrite (\ref{w11}) as a sum of two identical terms and to
interchange $\om$ and $\om'$ in one of these terms; this allows us to write
\ba
\label{w1-2}
W_1(\br,t)&=&{1\over 8\pi}\sum_{\lambda} \int_0^{\infty}\!d\om'\int_0^{\infty}\!d\om
\frac{\om'\eps_R(\om')-\om\eps_R(\om)}{\om'-\om} \la\bE_{\lambda}(\br,\om)\cdot\bE^{\dag}_{\lambda}(\br,\om')\ra 
e^{-i(\om-\om')t} \nonumber \\
&&
- {i\over 8\pi}\sum_{\lambda}\! \int_0^{\infty}\!\!\!\!d\om'\int_0^{\infty}\!\!\!\!d\om \left(\om'\eps_I(\om')+\om\eps_I(\om)\right) \nonumber \\
&& \times \mbox{}\frac{\la\bE_{\lambda}(\br,\om)\cdot\bE^{\dag}_{\lambda}(\br,\om')\ra e^{-i(\om-\om')t} - \la\bE_{\lambda}(\br,\om')\cdot\bE^{\dag}_{\lambda}(\br,\om)\ra e^{i(\om-\om')t}}{2(\om'-\om)}
\mbox{}+{1\over 8\pi}\la\bH^2(\br,t)\ra .\nonumber \\
\ea
It follows from (\ref{eqC}) and (\ref{eq33}) that the vacuum (zero temperature) expectation value
\ba
\la\bE_{\lambda}(\br,\om)\cdot\bE^{\dag}_{\lambda}(\br,\om')\ra = 
\la\bE_{\lambda}(\br,\om')\cdot\bE^{\dag}_{\lambda}(\br,\om)\ra 
= {\hbar\over 2\pi^3}\eps_I(\om){\om^4\over c^4} \int d^3k{1\over |k^2-
\eps(\om)\om^2/c^2|^2} \delta(\om-\om').
\ea
The first term in (\ref{w1-2}) is now evaluated using 
\ba
\lim_{\om'\rightarrow\om}{\om\eps_R(\om)-\om'\eps_R(\om')\over\om-\om'}={d\over d\om}[\om\eps_R(\om)].
\ea 
We evaluate the second term by noting that the zeroth-order contributions in $(\om-\om')$ in the numerator cancel each other, while the first-order terms result in a contribution linear in the elapsed time $t$:
\ba
\lim_{\om'\rightarrow\om}\frac{e^{-i(\om-\om')t} - e^{i(\om-\om')t}}{2(\om'-\om)}=it .
\ea
Therefore
\ba
W_1(\br,t)&=&\!{1\over 8\pi}{\hbar\over 2\pi^3c^4}\sum_{\lambda} \int_0^{\infty}\!\!\!d\om \!\left({d\over d\om}[\om\eps_R] + 2t\omega\eps_I \!\right) \!\om^4
\eps_I  \int d^3k{1\over |k^2-\eps\om^2/c^2|^2}+{1\over 8\pi}\la\bH^2(\br,t)\ra \nonumber \\
&=&{\hbar\over 8\pi^2c^3}\sum_{\lambda}\int_0^{\infty}\!\!\!d\om\om^3n_R {d\over d\om}[\om\eps_R] 
 +{1\over 8\pi}\la\bH^2(\br,t)\ra \, + \, t{\hbar\over 4\pi^2c^3}\sum_{\lambda}\int_0^{\infty}\!\!\!d\om\om^4 n_R \eps_I ,
\label{w1}
\ea
where we have used the relations $\eps_R=n_R^2-n_I^2$ and $\eps_I=2n_Rn_I$ between the real and imaginary parts of the
permittivity $\eps(\om)$ and the refractive index $n(\om)$. 

For the evaluation of $W_2$ it is convenient to define $\bK(\bk,\om)$ by writing
\be
\label{KTot}
\bK(\br,t)=\int_0^{\infty}d\om\int d^3k\sum_{\lambda}[\bK_{\lambda}(\bk,\om)e^{-i\om t}e^{i\bk\cdot\br}
+\bK^{\dag}_{\lambda}(\bk,\om)e^{i\om t}e^{-i\bk\cdot\br}],
\ee
and use (\ref{eqk}), (\ref{eqkk}), and (\ref{eq33}) to relate $\bK_{\lambda}(\bk,\om)$ and $\bE_{\lambda}(\bk,\om)$:
\be
\label{Klambda}
\bK_{\lambda}(\bk,\om)={c^2\over\om^2}[k^2-\eps(\om)\om^2/c^2]\bE_{\lambda}(\bk,\om).
\ee
After inserting (\ref{KTot}) and (\ref{Klambda}) in (\ref{W2}) and performing some algebra we get 

\ba
W_2(\br,t)&=&- \frac{\hbar}{16\pi^4 c^2} \sum_{\lambda} \int_0^{\infty}\!\!\! d\om' \int_0^{\infty} \!\!\! d\om \frac{\om^2\om'}{\om-\om'} \sqrt{\eps_I(\om)\eps_I(\om')} \delta(\om-\om') \nonumber \\
&& \times \mbox{}\int d^3k \left[ \frac{e^{-i(\om-\om')t}}{k^2-\eps(\om)\om^2/c^2} + \frac{e^{i(\om-\om')t}}{k^2-\eps^*(\om)\om^2/c^2} \right],
\ea
and, proceeding as in the evaluation of $W_1$, 
\ba
W_2(\br,t)&=& - {\hbar\over 8\pi^4 c^2}\sum_{\lambda}{\rm Re}\int_0^{\infty} \!\!\! d\om'\int_0^{\infty}\!\!\! d\om {\om^2\om'\over\om-\om'}\sqrt{\eps_I(\om)\eps_I(\om')} \delta(\om-\om')\nonumber \\
&&\mbox{}\times\int d^3k{1\over k^2-\eps(\om')\om'^2/c^2} -
t{\hbar\over 4\pi^2c^3}\sum_{\lambda}\int_0^{\infty}\!\!\!d\om\om^4 n_R \, \eps_I .
\label{w2-2}
\ea
We see that the time-dependent term in $W_2(t)$ exactly cancels the time-dependent term in $W_1(t)$.

The total energy density is obtained by adding (\ref{w1}) and (\ref{w2-2}) \cite{rosa}:
\ba
W&=&{\hbar\over 8\pi^2c^3}\sum_{\lambda}\int_0^{\infty}d\om\om^3 \bigg\{ {\rm Re}\left[n_R{d\over d\om}(\om\eps)+\eps^{3/2}\right]
+{1\over\om}\eps_I{\rm Im}{d\over d\om}(\om^2\eps^{1/2})\bigg\}.
\label{energyabsorption}
\ea
Finally, using $\eps(\om)=n^2(\om)$ and the following relations
\ba
n_R{d\over d\om}(\om\eps_R) &=& (n_R^2-n_I^2)n_R + \, \om n_R \left( 2n_R \frac{dn_R}{d\om} - 
2n_I \frac{dn_I}{d\om} \right) , \nonumber \\
{\rm Re} \, \eps^{3/2} &&\hspace{-8pt}= n_R^3 - 3 n_R n_I^2 , \nonumber \\
\frac{\eps_I}{\om} {\rm Im} \frac{d}{d\om} (\om^2 \sqrt{\eps}) &&\hspace{-8pt}= 4n_Rn_I^2 + 2n_Rn_I \om \frac{dn_I}{d\omega} ,
\ea
and summing over polarizations, we obtain the vacuum expectation value of the total energy density:
\ba
\label{FinalExpression}
W={\hbar\over 2\pi^2c^3}\int_0^{\infty}d\om\om^3n^2_R(\om)\left( n_R + \om {d n_R \over d\om} \right) 
={\hbar\over 2\pi^2c^3}\int_0^{\infty}d\om\om^3n^2_R(\om){d\over d\om}[\om n_R(\om)].
\ea
This has the exactly the form expected had we ignored absorption entirely and simply posited that
each mode of frequency $\om$ and wavenumber $k=n_R(\om)\om/c$ has a zero-point energy
$(1/2)\hbar\om$, so that the energy density summed over all modes is
\be
W=2\left({1\over 2\pi}\right)^3\int d^3k{1\over 2}\hbar\om.
\ee
The physical interpretation of this result is that the loss of energy due to
absorption is balanced by the work done by the Langevin forces that maintain the canonical commutation
relations of the dielectric oscillators.


\section{Summary}

Based on the simple model of a harmonic oscillator coupled to a reservoir, we showed how the Heisenberg
equations of motion leading to a Langevin equation for the oscillator can give results equivalent
to those obtained from the exact (Fano) diagonalization of the complete oscillator-reservoir system. We then
used the model of a dielectric medium as a collection of harmonic oscillators, each oscillator coupled to
a reservoir responsible for dissipation and a Langevin force as well as to the electromagnetic field, to derive the fluctuation-dissipation relation between the noise polarization arising from the Langevin forces and the imaginary part of the permittivity of the dielectric medium. 

The simple oscillator-reservoir model we considered would suggest that the solutions for the electric and magnetic fields in a dielectric medium, with the noise polarization as a source, might have the same form as obtained when the
complete system of dielectric oscillators, reservoirs, and the electromagnetic field is diagonalized. We showed
that this is in fact the case. Then we considered the {\sl total} energy density of the complete system and
showed explicitly that it is given by Eq. (\ref{FinalExpression}), which is exactly the form of the energy
density obtained when absorption is ignored. In particular, we showed that a positive energy rate $\dot{W}_1 > 0$ arising from the interaction of the electromagnetic field with the dielectric oscillators is exactly canceled by a corresponding negative energy rate coming from the interaction of the system with the reservoir, $\dot{W}_2 = -\dot{W}_1 < 0$.


\section*{Acknowledgements}

We thank S.M. Barnett, L.S. Brown, S.Y. Buhmann, I.E. Dzyaloshinskii, J.H. Eberly and R.F. O'Connell for helpful comments relating to this research. This work was funded by DARPA/MTO's Casimir Effect Enhancement program under DOE/NNSA Contract DE-AC52-06NA25396.



\begin{thebibliography}{99}
\bibitem{emenergy} T. Erber, ``Poynting Vector and Energy Density in Dispersive Media," Bull. Classe
des Sciences, Acad. Royale de Belgique {\bf 50}, 328 (1964).
\bibitem{sync} J. Schwinger, W.Y. Tsai, and T. Erber, ``Classical and Quantum Theory of Synergic 
Synchrotron-Cerenkov Radiation," Ann. Phys. (N.Y.) {\bf 96}, 303 (1976); {\bf 281}, 1019 (2000).
\bibitem{rr} T. Erber, ``The Classical Theories of Radiation Reaction," Fort. der Physik {\bf 9}, 
343 (1961).
\bibitem{erberqm} T. Erber, ``Testing the Randomness of Quantum Mechanics: Nature's Ultimate
Crypotgram?", Ann. N.Y. Acad. Sci. {\bf 755}, 748 (1995).
\bibitem{casimir} H.B.G. Casimir, ``On the Attraction between Two Perfectly Conducting Plates," Proc. K. Ned. Akad. 
Wet. {\bf 51}, 793 (1948).
\bibitem{vankampen} N.G. van Kampen, B.R.A. Nijboer, and K. Schram, ``On the Macroscopic Theory of van der
Waals Forces," Phys. Lett. {\bf 26A}, 307 (1968).
\bibitem{pwmbook} See, for instance, P.W. Milonni, {\sl The Quantum Vacuum. An Introduction to Quantum Electrodynamics}
(Academic, San Diego, 1994), and references therein.
\bibitem{ginzburg} V.L. Ginzburg, {\sl Theoretical Physics and Astrophysics} (Pergamon, Oxford, 1979), p. 309.
\bibitem{remark1} Correct results (in agreement with Lifshitz) are nevertheless obtained: an integral over frequency is 
analytically continued in such a way that the permittivity appears as a function of purely imaginary frequencies, at
which the {\sl complex} permittivity, like the permittivity in the absence of absorption, is purely real.
\bibitem{lifshitz} E.M. Lifshitz, ``The Theory of Molecular Attractive Forces between Solids," Sov. Phys. 
JETP {\bf 2}, 73 (1956). 
\bibitem{hutt} B. Huttner and S.M. Barnett, Phys. Rev. A {\bf 46}, 4306 (1992).
\bibitem{buhm} F.S.S. Rosa, S.Y. Buhmann, D.A.R. Dalvit, and P.W. Milonni, in preparation.
\bibitem{rosa} Details of some of the material presented here are provided in a preprint by F.S.S. Rosa, 
D.A.R. Dalvit, and P.W. Milonni, ``Electromagnetic Energy, Absorption, and Casimir Forces. I. Uniform 
Dielectric Media in Thermal Equilibrium.", arXiv:0911.2736.
\bibitem{fano} U. Fano, ``Atomic Theory of Electromagnetic Interactions in Dense Materials," 
Phys. Rev. {\bf 103}, 1202 (1956).
\bibitem{remark2} See Reference \cite{rosa} for a discussion of the relation between the form of fluctuation-dissipation relation obtained here and that in Lifshitz's paper.
\end{thebibliography}
\end{document}